\documentstyle[aps,prb,twocolumn,graphics]{revtex}

\begin{document}

\title{The probability distribution of the conductance at the mobility edge}
\author{Marc R{\"{u}}hl{\"{a}}nder$^1$ and C. M. Soukoulis$^{1,2}$\\
$^1$Ames Laboratory and Department of Physics and Astronomy\\
Iowa State University, Ames, Iowa 50012\\
$^2$Research Center of Crete and Department of Physics\\
%University of Crete, Heraklion, Crete, Greece}
University of Crete, Heraklion, Crete, Greece\\[0.25in]
\begin{minipage}{5.3in}
The probability distribution of the conductance $p(g)$ of disordered
2d and 3d systems is calculated by transfer matrix techniques. As
expected, $p(g)$ is Gaussian for extended states while for localized
states it is log--normal. We find that at the mobility edge $p(g)$ is
highly asymmetric and universal.
\end{minipage}}
\maketitle

%\begin{abstract}
%The probability distribution of the conductance $p(g)$ of disordered
%2d and 3d systems is calculated by transfer matrix techniques. As
%expected, $p(g)$ is Gaussian for extended states while for localized
%states it is log--normal. We find that at the mobility edge $p(g)$ is
%highly asymmetric and universal.
%\end{abstract}

In the presence of disorder \cite{Sou99b}, a system may undergo a transition 
from insulating to metallic behaviour as the Fermi energy varies in an
energy range containing both localized and extended states, seperated
by a mobility edge. This transition can be characterized either by
transport properties like e.g.\ the conductance, or by properties of
the eigenstates of the system like e.g.\ the correlation length $\xi_c$
(approaching from the metallic side of the transition) or the localization
length $\xi_l$ (approaching from the insulating side of the transition).
While the latter are self--averaging quantities, i.e.\ the ensemble
average may be used as a scaling variable, the conductance is not
\cite{Sle97,Sou99a,Mks99,Mtt99,Wan98b,Ple98}.
Therefore it is of great importance to determine the complete probability
distribution $p(g)$ of the conductance $g$
(in units of $e^2/h$), especially
at the critical point of the metal--insulator transition, as it is well
known to be a Gaussian on the metallic side and log--normal on the
insulating side. The correct form of $p_c(g)$, the distribution of the
conductance at the mobility edge
\cite{Sle97,Sou99a,Mks99,Mtt99,Wan98b,Ple98}, 
is still not sufficiently well--known.

\begin{figure}[h]
\resizebox{3.0in}{3.0in}{ \includegraphics{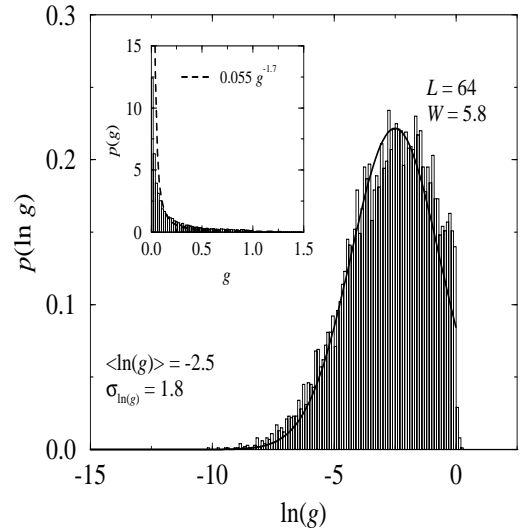} }
\caption{\label{crit2da0} The distribution of the conductance for a square 
system of 64 by 64 lattice sites. The disorder was chosen such that the
localization length is close to the system size $L$.}
\end{figure}

\begin{figure}[h]
\resizebox{3.0in}{3.0in}{ \includegraphics{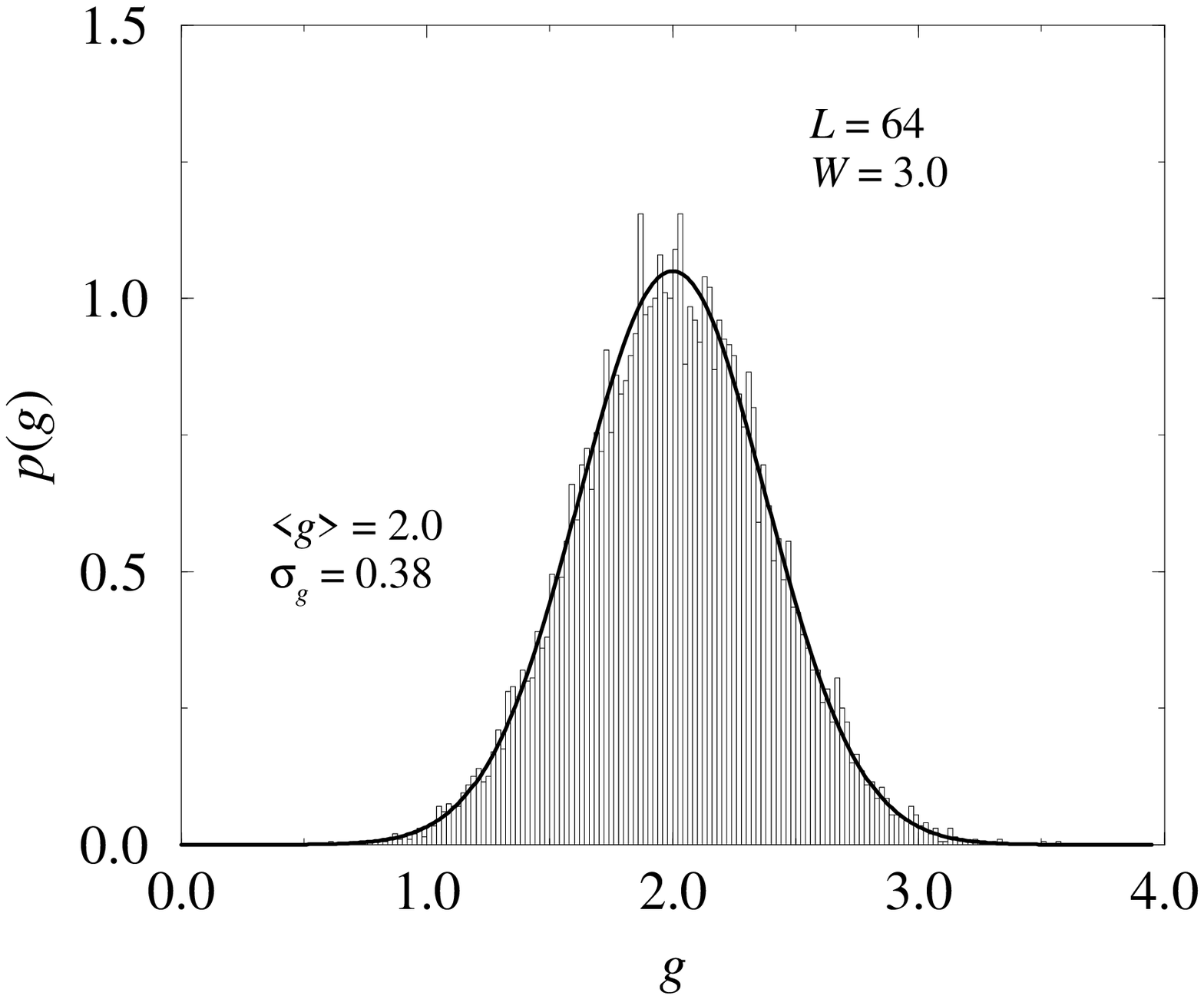} }
\resizebox{3.0in}{3.0in}{ \includegraphics{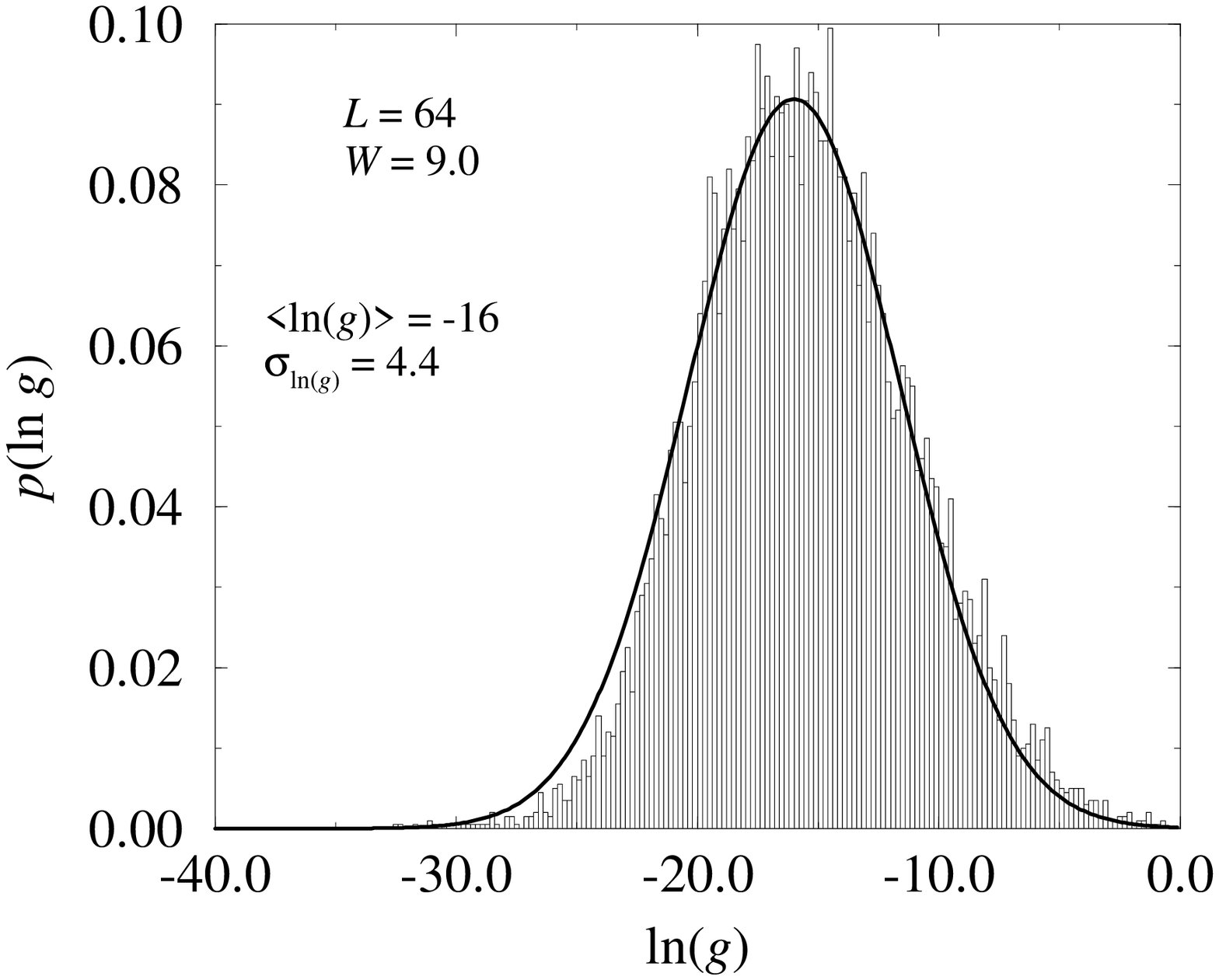} }
\caption{\label{non-crit2da0} The distribution of the conductance for a square
of 64 by 64 lattice sites. Small disorder ($W = 3$, top panel) leads to
metallic behaviour; strong disorder ($W = 9$, bottom panel) leads to
insulating behaviour. The values for the average and standard deviation
of the Gaussian fits are given in the graphs.}
\end{figure}

Using a tight--binding model \cite{Eco83} with diagonal disorder, a 
transition from a metallic state to an insulating one can be induced 
\cite{Krm93} in a finite size
sample by increasing the disorder strength $W$. In all our results,
$W$ is given in units of the hopping integral. The localization length
$\xi_l$ decreases as the strength of the disorder, $W$, increases. 
As long as $\xi_l$ is much bigger
than the system size, the electron will cross the sample with ease, thus
being essentially delocalized. If, on the other hand, $\xi_l$ is sufficiently
smaller than the system size, the electron will become localized in a small
region and not contribute much to the conductance. The critical strength
of disorder $W_c$ will occur where the localization length becomes
comparable to the system size. 
We have systematically studied the conductance $g$ of the 2d and 3d 
tight--binding model by using the transfer matrix technique, which
relates $g$ with the transmission matrix ${\bf t}$ by $g = 2 {\rm Tr}
({\bf t}^\dagger {\bf t})$. The $g$ defined here is for both spin
orientations.
Fig.\ \ref{non-crit2da0} shows the
distributions of the conductance $g$ and the natural logarithm of the
conductance $\ln(g)$ for a metallic and a localized 2d sample respectively.
In the ``metallic'' regime (it is really weakly localized, since we are
working in 2d) where $W = 3.0$ we have $\xi_l = 47234$, which is much larger
than the system size $L = 64$. In the localized regime, where $W = 9.0$,
we have $\xi_l = 7.54$, which is smaller than the size of the system.
Both distributions can be fitted very well by a Gaussian normal distribution
\cite{Ple98}.
Fig.\ \ref{crit2da0} displays the distributions for a system near the
critical point. The distribution of $\ln(g)$ shows a characteristic
cut--off at $\ln(g) = 0$. Only a few samples have a conductance $g > 1$.
This is in agreement with analytical results \cite{Mtt99}.

\begin{figure}[h]
\resizebox{3.0in}{3.0in}{ \includegraphics{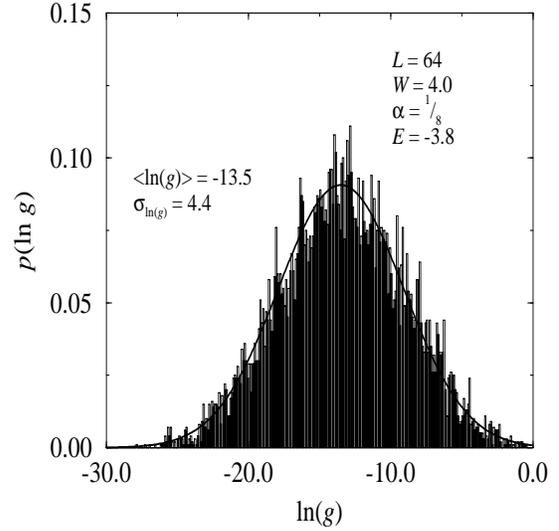} }
\caption{\label{non-crit2da8} The conductance distribution of a square 
system of 64 by 64 lattice sites. For the chosen disorder strength $W = 4$
and magnetic flux $\alpha = 1/8$ states at the energy $E = -3.8$ are well
localized. The parameters for the Gaussian fit are given in the graph.}
\end{figure}

\begin{figure}[h]
\resizebox{3.0in}{3.0in}{ \includegraphics{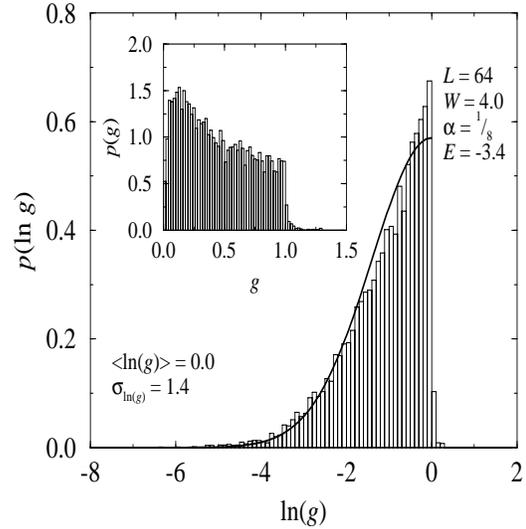} }
\caption{\label{crit2da8} The conductance distribution of a square system
of 64 by 64 lattice sites. For the chosen disorder strength $W = 4$ and
magnetic flux $\alpha = 1/8$ states at the energy $E = -3.4$ are critical.
The parameters for the fit are given in the graph.}
\end{figure}

Applying a strong magnetic field perpendicular to a 2d system creates states
with a diverging localization length \cite{Wan98b} even in the thermodynamical 
limit, as long as the disorder is not too strong. Thus, one can approach a
critical state in such a system by varying the energy, even though one
cannot reach a true metallic state. In Fig.\ \ref{non-crit2da8} we have
an insulating system, the conductance distribution again fitting well to
a log--normal distribution. The distribution for the critical state $p_c(g)$,
shown in Fig.\ \ref{crit2da8}, again has the abrupt cut--off at $\ln(g) = 0$.
It can be fitted to a skewed log--normal distribution, which is normalized
on the interval $(-\infty;0]$ rather than all real numbers.
Notice that in this case too $p_c(g)$ is highly asymmetric and very similar
to the case shown in Fig.\ \ref{crit2da0}.

\begin{figure}[h]
\resizebox{3.0in}{3.0in}{ \includegraphics{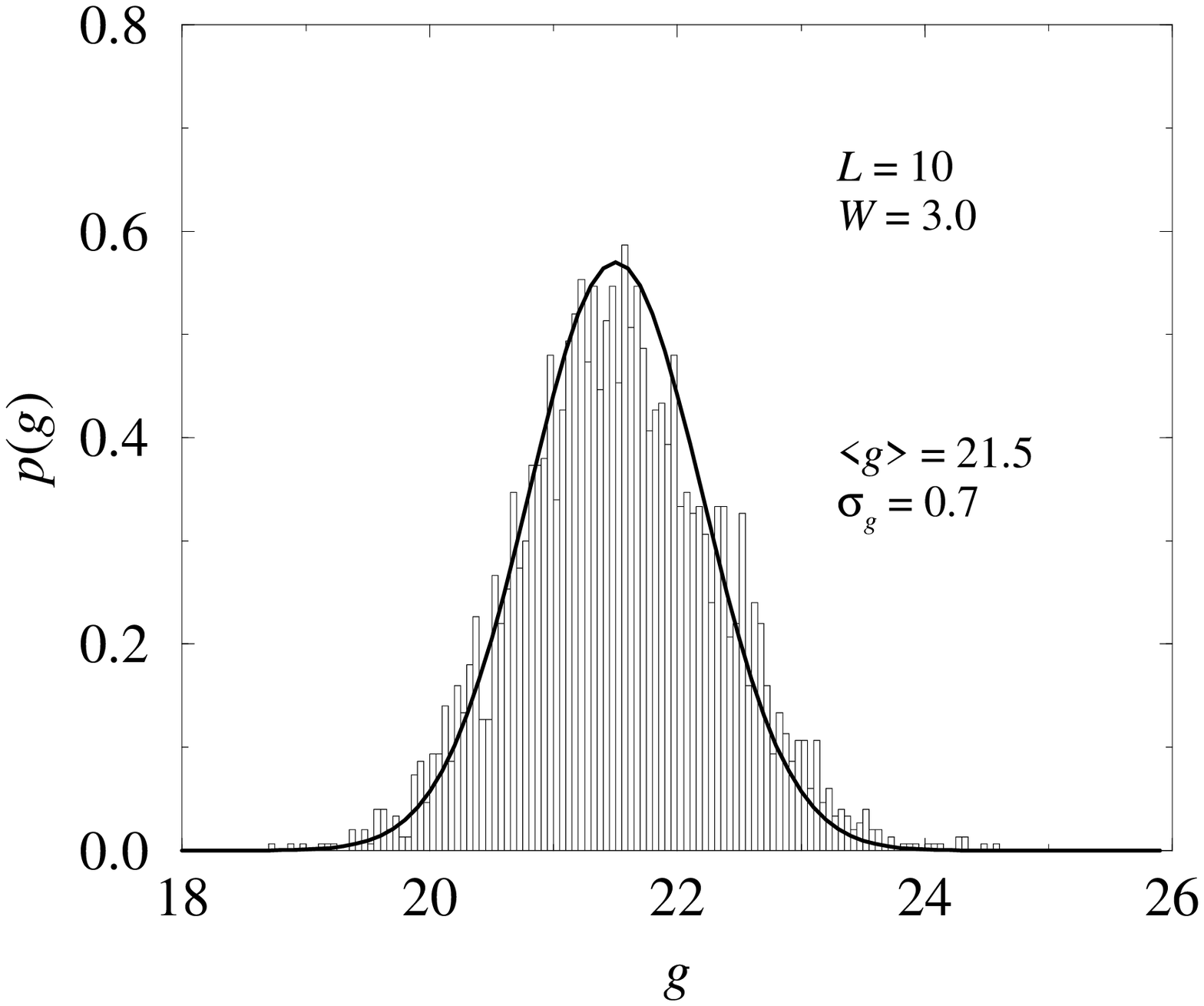} }
\resizebox{3.0in}{3.0in}{ \includegraphics{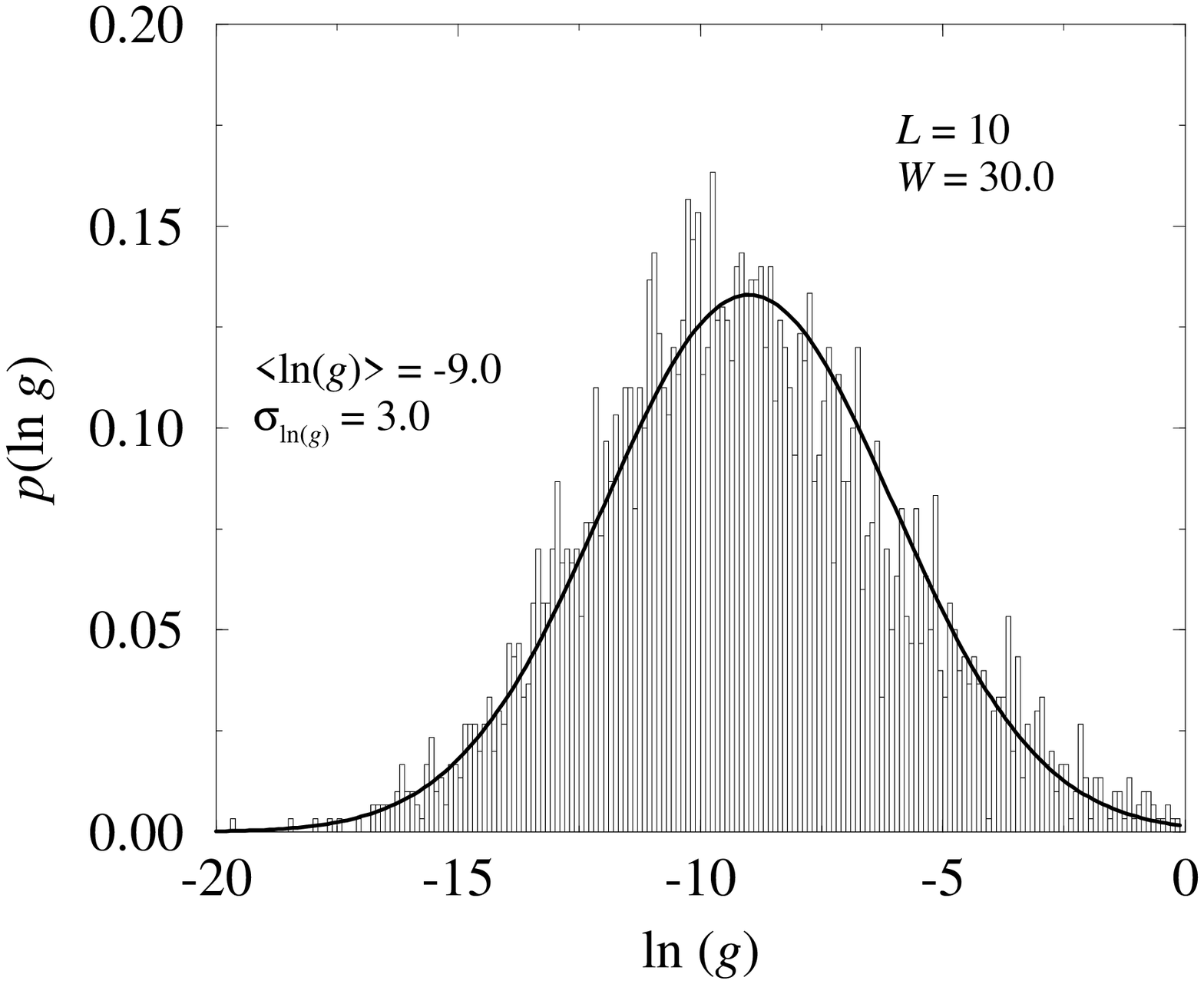} }
\caption{\label{non-crit3d} The conductance distribution of a cubic system
of 10 by 10 by 10 lattice sites. Small disorder ($W = 3$, left panel) leads 
to metallic behaviour; strong disorder ($W = 30$, right panel) leads to
insulating behaviour. The values for the average and standard deviation of
the Gaussian fits are given in the graphs.}
\end{figure}

\begin{figure}[h]
\resizebox{3.0in}{3.0in}{ \includegraphics{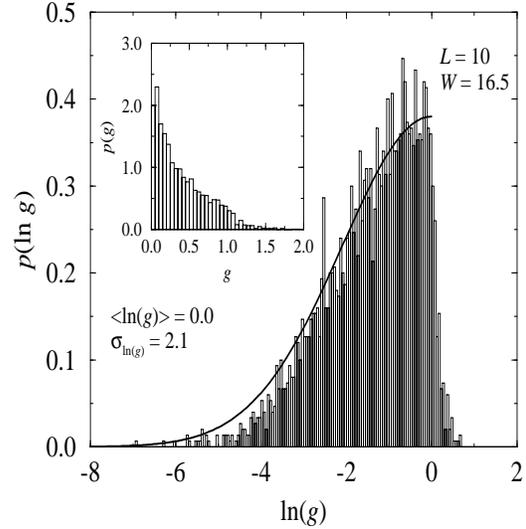} }
\caption{\label{crit3d} The conductance distribution of a cubic system of
10 by 10 by 10 lattice sites. The disorder was chosen such that the
localization length is close to the system size $L$. The parameters for
the fit are given in the graph.}
\end{figure}

Comparing these results to a 3d system without magnetic field, again
varying the disorder strength and keeping the energy fixed at $E = 0.0$, 
we find the
same qualitative picture: the distribution of the conductance is normal
on the metallic side and log--normal on the insulating side of the
transition (see Fig.\ \ref{non-crit3d}), whereas the critical state is
characterized by a cut--off at $\ln(g) = 0$ and a skewed log--normal
distribution for $\ln(g) \leq 0$ (see Fig.\ \ref{crit3d}).

In conclusion, our detailed numerical results show that the probability
distribution of the conductance is normal for the extended regime and
log--normal for the localized regime. However, at the mobility edge
$p(g)$ is highly asymmetric. The form of $p_c(g)$ at the critical point
is independent of the dimensionality of the system and of the model.
This suggests that $p_c(g)$ is universal.

Ames Laboratory is operated for the U.S.\ Department of Energy by
Iowa State University under Contract No.\ W--7405--Eng--82.

\end{document}